\documentclass[superscriptaddress,twocolumn]{revtex4}
\usepackage{mathrsfs}
\usepackage{amssymb}
\usepackage[tbtags]{amsmath}
\usepackage{graphicx}
\usepackage{epsfig,graphicx,times}
\usepackage{graphicx}
\usepackage{subfigure}
\usepackage{color}
 \usepackage[titletoc]{appendix}

\begin{document}
\title{Experimental evidences of a current-biased Josephson junction device can be worked as a macroscopic ``Boson" or ``Fermion" and the combination}
\author{P. H. Ouyang}
\affiliation{Information Quantum Technology Laboratory, International Cooperation Research Center of China Communication and Sensor Networks for Modern Transportation, School of Information Science and Technology, Southwest Jiaotong University, Chengdu 610031, China}
\author{S. R. He}
\affiliation{Information Quantum Technology Laboratory, International Cooperation Research Center of China Communication and Sensor Networks for Modern Transportation, School of Information Science and Technology, Southwest Jiaotong University, Chengdu 610031, China}
\author{Y. Z. Wang}
\affiliation{Information Quantum Technology Laboratory, International Cooperation Research Center of China Communication and Sensor Networks for Modern Transportation, School of Information Science and Technology, Southwest Jiaotong University, Chengdu 610031, China}
\author{Y. Q. Chai}
\affiliation{Information Quantum Technology Laboratory, International Cooperation Research Center of China Communication and Sensor Networks for Modern Transportation, School of Information Science and Technology, Southwest Jiaotong University, Chengdu 610031, China}
\author{J. X. He}
\affiliation{Information Quantum Technology Laboratory, International Cooperation Research Center of China Communication and Sensor Networks for Modern Transportation, School of Information Science and Technology, Southwest Jiaotong University, Chengdu 610031, China}
\author{H. Chang}
\affiliation{Qixing Vacuum Coating Technology Co., Ltd., Chengdu 610031, China}
\author{L. F. Wei\footnote{Correspondence author: lfwei@swjtu.edu.cn}}
\affiliation{Information Quantum Technology Laboratory, International Cooperation Research Center of China Communication and Sensor Networks for Modern Transportation, School of Information Science and Technology, Southwest Jiaotong University, Chengdu 610031, China}

\date{\today}
\begin{abstract}
According to the statistical distribution laws, all the elementary particles in the real 3+1-dimensional world must and only be chosen as either bosons or fermions, without exception and not both. Here, we experimentally verified that a quantized current-biased Josephson junction (CBJJ), as an artificial macroscopic ``particle", can be served as either boson or fermion, depending on its biased dc-current. By using the high vacuum two-angle electron beam evaporations, we fabricated the CBJJ devices and calibrated their physical parameters by applying low-frequency signal drivings. The microwave transmission characteristics of the fabricated CBJJ devices are analyzed by using the input-output theory and measured at $50$mK temperature environment under low power limit. The experimental results verify the theoretical predictions, i.e., when the bias current is significantly lower than the critical one of the junction, the device works in a well linear regime and thus works as a harmonic oscillator, i.e., a ``boson"; while if the biased current is sufficiently large (especially approaches to its critical current), the device works manifestly in the nonlinear regime and thus can be served as a two-level artificial atom, i.e., a ``fermion". Therefore, by adjusting the biased dc-current, the CBJJ device can be effectively switched from the boson-type macroscopic particle to the fermion-type one, and thus may open the new approach of the superconducting quantum device application.
\end{abstract}
\maketitle

\section{Introduction}
Spin quantum statistics, related to the well-known Pauli exclusion principle, is one of the basic features in quantum mechanics ~\cite{Zee}. In three-dimensional space, the microscopic particles with either integer or half-integer spin quantum numbers and thus can be divided into the bosons and fermions, respectively. Phenomenologically, the boson-type particles (such as photons and phonons) obey the Bose-Einstein statistics and the relevant bosonic creation and annihilation operators; $\hat{a}^\dagger$ and $\hat{a}$, satisfy the bosonic communication relation: $[\hat{a},\hat{a}^\dagger]=\hat{a}\hat{a}^\dagger-\hat{a}^\dagger\hat{a}=1$; while the fermion-type particle (typically such as the electrons) obey alternatively the Fermi-Dirac statistics, whose the creation and annihilation operators; $\hat{b}^\dagger$ and $\hat{b}$; satisfy alternatively the anti-communication relation: $\{\hat{b},\hat{b}^\dagger\}_+=\hat{b}\hat{b}^\dagger+\hat{b}^\dagger\hat{b}=1$.
Basically, all objects called the elementary particles in the real 3+1-dimensional world must and only be chosen as either bosons or fermions, without exception and not both.

However, in two-dimensional space certain ``particles" can lie somewhere between the bosons or fermions~\cite{science1990}, say anyons. For example, the elementary excitations of the fractional quantum Hall effect systems at filling factor $\nu=1/m$ (with $m$ being an odd integer) have been predicted to obey Abelian fractional statistics, which have detected recently by measuring the current correlations of the collision between the anyons (with the filling factor: $\nu=1/3$) at a beamsplitter~\cite{science2020}.
While, the non-Abelian anyons~\cite{RMP} which had just been verified indirectly~\cite{nature2018-1,nature2018-2},
are particularly expected for the desired fault-tolerant quantum computation~\cite{Kitaev2003}.
A natural question is, if the statistical behaviors for the three-dimensional macroscopic artificial objects, working in different parameter regimes, could still be controllable? If it is true, then the different applications basing on their mutually-exclusive statistical features could be realized with the same devices. For example, if certain elementary devices could be manipulated as either the bosons sometimes or the fermion at the other times, then their controllability could be significantly improved in the compacted solid-state circuits on chip~\cite{Google}.

In fact, a single bosonic Josephson junction, implemented by two weakly
linked Bose-Einstein condensates in a double-well potential, had been experimentally realized~\cite{PRL2005}. Also, it has been shown that under the low-excitation limit, a Josephson junction with the small dc-current bias could be effectively treated as a linear harmonic oscillator~\cite{PRB2005}, which can quantized as a simple-mode boson and thus could be severed as the quantum data bus to implement the couplings between the distant superconducting qubits. While, if the biased dc-current is sufficiently large (e.g., approaching to its critical current), the quantized CBJJ device has only a few  bound states~\cite{Martinis-science} and thus can be treated as a macroscopic qubit (i.e., a fermion) called as the CBJJ one~\cite{Martinis-PRL} for the superconducting quantum information computing~\cite{science2003}. This implies that, a current-biased Josephson junction (CBJJ), acted as one of the artificially macroscopic quantum system, can be served as either the boson or the fermion, depending on its biased current.

In this work we provide directly the experimental evidences, i.e., a CBJJ device is really a combinator of the macroscopic boson- and fermion object, i.e., it can be worked as either a boson or a fermion under the extreme dc-current biases, and both of them for the other biased regime.
First, by using the standard input-output theory, we predicate the microwave scattering features of the device with the two extreme biases; one is to generate a boson, and the other is to generate a fermion. Next, the device is fabricated and then measured under the low-power microwave drivings at ultra-low temperature. The experimental results agree well with the corresponding theoretical predictions, i.e., the quantized CBJJ device can be served as either a macroscopic boson for the relatively low dc-current bias or a macroscopic fermion under the sufficiently large dc-current bias, under the low-excitation limit. Between these biases, the device works the boson with certain probabilities and also the fermion with the other ones. With these theoretical predications and also the experimental verifications, a novel superconducting circuit to implement the macroscopic quantum information processing might be constructed, wherein the quantized CBJJ devices play the double roles; their fermonic behaviors can be utilized to encode the CBJJ qubits, and their bonsonic features can be severed as the data buses to implement the indirectly coupling between the distant qubits and the high fidelity readouts of the solid-state qubits.

\section{Models and Experimental Predictions}

We first develop a theoretical approach to confirm a quantized CBJJ device, under the different dc-current biases, can be served as either a macroscopic boson or a macroscopic fermion. These predictions will be verified experimentally later.

\subsection{Macroscopic quantum bound states of a current-biased Josephson Junction}

As a macroscopic quantum device, the phase $\delta$ dynamics of a CBJJ shown in Fig.~1, is equivalent to that of a particle with the mass $C_J[\Phi_0/(2 \pi)]^2$ (with $C_J$ being the effective capacitance of the junction) moving in a potential~\cite{Martinis-science, Blais-PRL}:
$$
U(\delta)=\frac{I_{c} \Phi_0}{2 \pi}\left(-\frac{I_b}{I_{c}} \delta-\cos \delta\right)
=-E_J\left(\frac{I_b}{I_{c}} \delta+\cos \delta\right),
$$
where $\Phi_{0}=h/(2e)=2.07 \times 10^{-15}$Wb is the flux quanta, $\delta=2 \pi \phi/\Phi_{0}$ is the gauge invariant phase across the junction, $\phi$ is the flux, $I_b(<I_c)$ is the biased dc-current; $I_c$ and $E_J=2\pi I_c/\Phi_0$ are the critical current and Josephson energy of the junction, respectively.
	\begin{figure}[h]
		\centering
		\includegraphics[width=5cm]{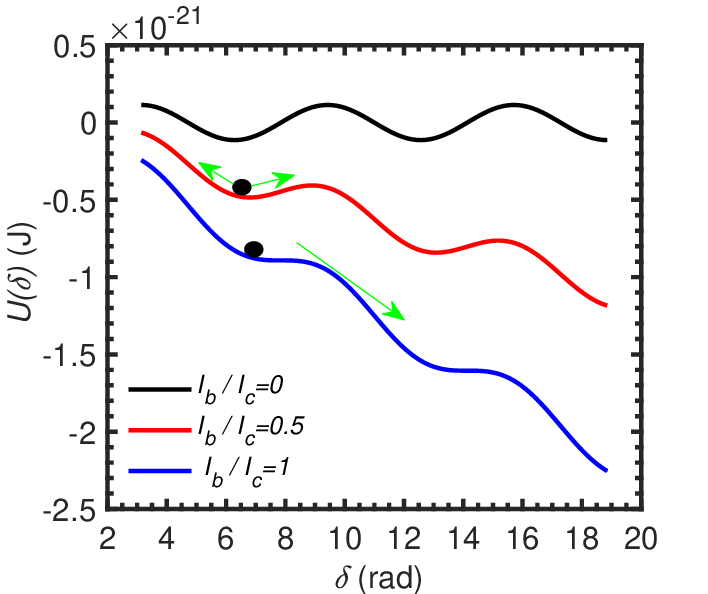}
		\caption{Under the dc-biased current $I_b<I_c$, the CBJJ is equivalent to a macroscopic phase particle with the mass $m$ moving in a wash potential, which contains a series of potential traps with the limited heights $\Delta U=U_{\rm{max}}-U_{\rm{min}}$. Here, the black ball refers to the phase particle, which moves in the trap with the plasma frequency $\omega_a$, and the junction parameters are set as: $I_c=0.975{\rm{~\mu A}}, C_J=93{\rm{~fF}}$.}
		\label{Fig1:1}
	\end{figure}
It is seen that, this wash-potential contains a series of potential traps, whose lowest- and highest potentials are determined by
$$
\frac{dU(\delta)}{d\delta}=-E_J\left(\frac{I_b}{I_c}-\sin\delta\right)|_{\delta
=\delta_{\rm{min}},\delta_{\rm{max}}}=0,
$$
with $d^2U(\delta)/d\delta^2|_{\delta=\delta_{\rm{min}}}>0$ and $d^2U(\delta)/d\delta^2|_{\delta=\delta_{\rm{max}}}<0$, respectively.

It is easily proved that:
$$
\begin{aligned}
\delta_{\rm{min}}=2k\pi+\sin^{-1}\left(\frac{I_b}{I_c}\right),\,\delta_{\rm{max}}=3k\pi-\sin^{-1}\left(\frac{I_b}{I_c}\right),
\end{aligned}
$$
with $k=0,1,2......$. For example, between $\delta=\pi$ and $\delta=3\pi$, we have the local minimal potential: $U_{\rm{min}}$ at $\delta_{\min }=2\pi+\sin^{-1}(I_b/I_c)$ and the local maximal one: $U_{\rm{max}}$ at $\delta_{\min }=3\pi+\sin^{-1}(I_b/I_c)$. Of course, if $I_b\geq I_{c}$, there is not any quantum bound state in the trap and thus the Josephson behaves a resistor of $R_N$, showing as a non-zero voltage state: $V=I_bR_N$.
Near the bottom of the potential trap, i.e., $\delta\sim\delta_{\rm{min}}$, the local potential of the phase particle can be expressed as
\begin{eqnarray}
U(\delta)&=&U(\delta_{\rm{min}})+\frac{1}{2}\frac{d^2U(\delta)}{d\delta^2}(\delta-\delta_{\rm{min}})^2
+...\nonumber\\
&\approx&\frac{1}{2}m\omega^2_p(\delta-\delta_{\rm{min}})^2,
\end{eqnarray}
with the plasma frequency~\cite{PR-2016}:
$$
\omega_p(I_b)=\omega_{p}\sqrt[4]{1-\left(\frac{I_b}{I_{c_b}}\right)^2},
$$
where $\omega_{p}$ is the plasma frequency for $I_b=0$. Mathematically, the height of the local potential trap can be calculated as:
\begin{eqnarray}
\Delta U&=&U(\delta_{\rm {\rm{max}}})-U(\delta_{\rm {\rm{min}}})\\
&=&2E_J\left(\sqrt{1-\left(\frac{I_b}{I_{c_b}}\right)^2}
-\left(\frac{I_b}{I_{c_b}}\right)\cos ^{-1}\left(\frac{I_b}{I_{c_b}}\right)\right),\nonumber
\end{eqnarray}
which is controllable by adjusting the amplitude of the biased dc-current $I_b$.

At ultra-low temperatures, the quantized CBJJ device can be described by a junction capacitance $C_J$ paralleled to a nonlinear inductance $L_J$, and thus it can be described by a quantized Hamiltonian~\cite{Martinis-science}:
\begin{equation}
\hat{H}_{CBJJ}=\frac{\hat{p}_\delta^2}{2 C_J\left(\frac{\Phi_0}{2\pi}\right)^2}+U(\delta).
\end{equation}
The macroscopic bound states with a series of discrete levels can be obtained by solving the following stationary Schr\"odinger equation:
$$
\hat{H}_{CBJJ}|n\rangle=E_n|n\rangle, n=0,1,2,...
$$
Specifically, for $C_J=93{\mathrm{~fF}}$ table 1 lists the corresponding transition frequencies and the "dipole-transition" matrix elements between the neighbour stationary states with the lower energies, under the typical biased currents, e.g., $I_b/I_c=0, 0.45, 0.68, 0.945, 0.955, 0.964$ and $0.97$, respectively.
\begin{table}[h!]
\centering
\caption{Energy differences between the nearest-neighbor levels of the lower bound states $|n\rangle$ of the CBJJ and their ``dipole-transition" matrix elements $\delta_{nm}=\langle n|m\rangle$ for the typical biased relative currents $I_b/I_c$. Here, $C_J=93{\mathrm{~fF}}$, $\omega_{mn}=(E_m-E_n)/\hbar,\,n\neq m=0,1,2,3$, and the unit of the energy is $10^{-24}{\rm{~J}}$.}
\begin{tabular}{c  c  c  c  c  c c}
  \hline
  $n$ & $\omega_{01}$ & $\omega_{12}$ & $\omega_{23}$ & $\delta_{01}$ & $\delta_{12}$ & $\delta_{23}$ \\
  \hline
  $0.000$ & $1.720$ & $1.719$ & $1.719$ & $0.052$ & $0.073$ & $0.089$ \\
  $0.450$ & $1.717$ & $1.716$ & $1.716$ & $0.055$ & $0.077$ & $0.094$ \\
  $0.680$ & $1.708$ & $1.704$ & $1.702$ & $0.060$ & $0.085$ & $0.104$ \\
  $0.945$ & $1.661$ & $1.593$ & $1.506$ & $0.091$ & $0.131$ & $0.163$ \\
  $0.955$ & $1.631$ & $1.526$ & $1.303$ & $0.096$ & $0.139$ & $0.161$ \\
  $0.964$ & $1.616$ & $1.450$ &         & $0.103$ & $0.151$ &         \\
  $0.970$ & $1.602$ &         &         & $0.108$ &         &         \\
  \hline
\end{tabular}
\label{table:1}
\end{table}
It is seen that, for the sufficiently low biased current, the energy differences between the lower macroscopic bound states in the potential trap are almost the same, with almost the equivalent ``dipole-transition" probabilities.
While, for the sufficiently large biased currents, the number of the macroscopic bound states in the potential trap are very limited and the relevant energy difference, as well as the ``dipole-transition" probabilities, between them are significantly different. Specifically, Fig.~2 shows how these behaviors changes with the biased dc-current.
\begin{figure}[h]
\centering
\includegraphics[width=5cm]{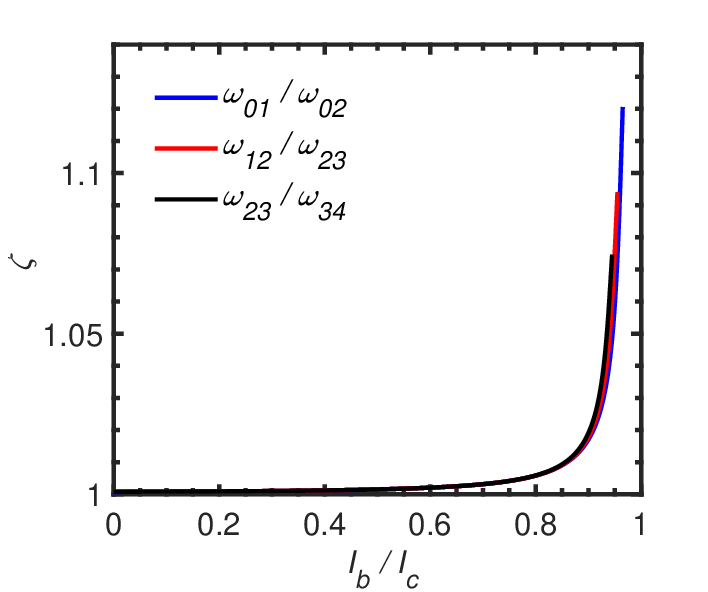}
\caption{The ratio $\zeta=\omega_{nm}/\omega_{mk},\, n=0,1,2,\,m=n+1, k=m+1$, of the transition frequencies between the nearest-neighbor bound states of the CBJJ versus the biased current. Here, the relevant parameters are set as: $I_c=0.975{\rm{~\mu A}}, C_J=93{\rm{~fF}}$. Noted that here just three, four, and five bound states are allowed for $I_b/I_{c}\leqslant 0.965, 0.957$, and $0.951$, respectively.}
\label{Fig3:3}
\end{figure}

Basing on the above numerical results, one can argue that, under the low-power driving, the CBJJ with the sufficiently low bias current can work as a linear harmonic oscillator, whose coherence time could be sufficiently long (due to the relatively low biased current noise).
In this case, the Hamiltonian of the quantized CBJJ device can be expressed as~\cite{PRB2005}:
\begin{equation}
\hat{H}_{b}\approx\frac{1}{2m_J}\hat{p}_{\theta}^{2}+\frac{1}{2}m_J\omega_p^2(I_b)\hat{\theta}^2
=\hbar\omega_p(I_b)\left(\hat{a}_p^\dagger\hat{a}_p+\frac{1}{2}\right),
\end{equation}
where $\hat{\theta}=\hat{\delta}-\delta_{\rm{min}}$,
$\hat{a}_p$ and $\hat{a}_p^\dagger$
are the $I_b$-dependent bosonic annihilation and creation operators, respectively. While, if the biased dc-current is sufficiently large, which yields the significantly strong nonlinearity, then the quantized CBJJ device should be served as an artificial atom with the transition between the bound states possessing the well selectivity, due to the manifest difference between the transition frequencies. For example, under the limit: $I_b\lesssim I_c$, the lowest two levels of the quantized CBJJ device could be utilized to encode a CBJJ qubit with the Hamiltonian~\cite{Martinis-PRL}:
\begin{equation}
H_{f}\approx \frac{\hbar \omega_{01}(I_b)}{2}\sigma_z,
\end{equation}
where $\sigma_z=|1\rangle\langle 1|-|0\rangle\langle 0|$ is the Pauli-type fermionic operator, satisfying the communication relation: $[\sigma_z,\sigma_{\pm}]=\pm2\sigma_{\pm},\,[\sigma_{+},\sigma_{-}]=\sigma_z$, with $\sigma_+=|1\rangle\langle 0|$ and $\sigma_-=|0\rangle\langle 1|$.

Obviously, under the low-excitation limit, the quantized CBJJ device can be worked as either a macroscopic boson or a macroscopic fermion, depending on its different dc-current biases for the controllable eigenfrequency. This is a novel feature for a macroscopic object, and thus it is desired to be tested both theoretically and experimentally.

\subsection{Transport properties of the low-power traveling-wave photons scattered by the CBJJ device under different biases}

To theoretically confirm the above argument that a quantized CBJJ can be served as either a boson or a fermion, let us investigate the scattering properties of the traveling-wave photon transporting along a transmission line.

\begin{figure}[h]
\centering
\includegraphics[width=4cm]{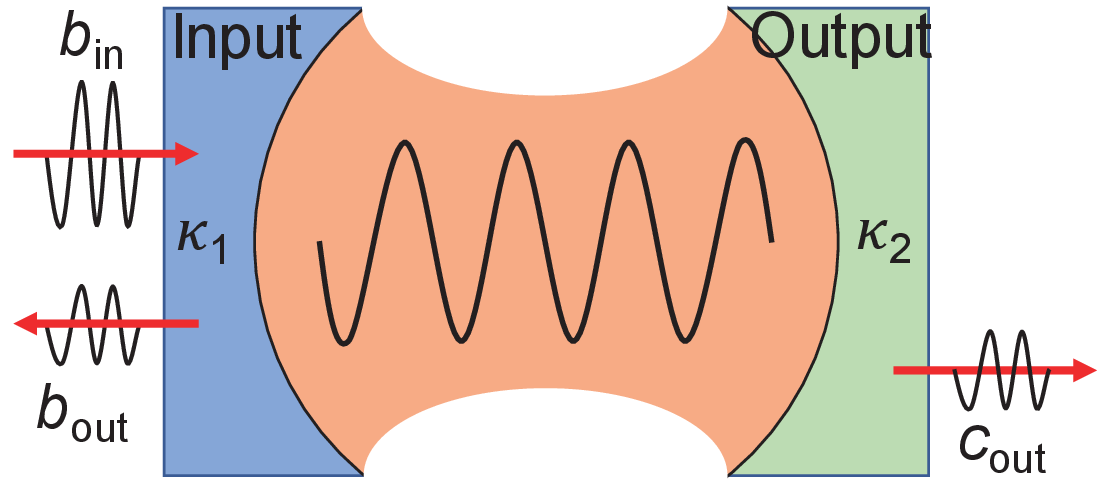}
\caption{A configuration of the traveling-wave photons scattered by the CBJJ, worked as a macroscopic boson (i.e., a simple-mode cavity). Here, $\hat{b}_{\rm{in}}$ and $\hat{b}_{\rm{out}}$ are the amplitude of the photon in/out of the CBJJ, and $\hat{c}_{\rm{out}}$ the one of the right/side photon}
\label{Fig4:4}
\end{figure}

First, if the CBJJ serves as a boson, which is physically equivalent to a quantized bosonic mode of a cavity, then the configuration shown in Fig.~\ref{Fig4:4} can be utilized to describe the transport properties of the traveling-wave photons scattered by the CBJJ.
The Hamiltonian of the present system can be expressed as: $\hat{H}_B=\sum_{l=b,c}\hat{H}_l+H_{CBJJ}+\hat{H}_{CBJJ-TL}$, where
\begin{equation}
\hat{H}_l=\int d\omega\sum_{l=b,c}\hbar\omega \hat{l}^\dagger(\omega)\hat{l}(\omega),
\end{equation}
describes the traveling-wave microwave photon transporting along the transmission line with $l=b, c$ refering to its left, right side, and the relevant bosonic operators satisfy the communication relation: $[l(\omega),l^{\dagger}(\omega^{\prime})]=\delta(\omega-\omega^{\prime})$. Also, the flux operator of the traveling-wave photon reads~\cite{Blais-PRL,PR-2016,Collett}:
\begin{equation}
\hat{\phi}_{l}(x)=\sqrt{\frac{\hbar Z_0}{4\pi}}\int_{0}^\infty\frac{d\omega}{\sqrt{\omega}}
\left[\hat{l}(\omega)e^{ikx}+\hat{l}^\dagger(\omega)e^{-ikx}\right],
\end{equation}
with $Z_0$ being the characteristic impedance of the transmission line, and thus
\begin{eqnarray}
\dot{\hat{\phi}}_l(x)=(-i)\sqrt{\frac{\hbar Z_0}{4\pi}}\int^{\infty}_0
d\omega\sqrt{\omega}\left[\hat{l}(\omega)e^{ikx}-\hat{l}^\dagger(\omega)e^{-ikx}\right].
\end{eqnarray}
Under the sufficiently low current bias, the CBJJ Hamiltonian reads:
$
\hat{H}_{CBJJ}\approx \hat{H}_{b},
$
shown in Eq.~4. The physical boundary condition at $x=0$, i.e., the location of the device, reads: $\hat{I}\left(0_b,t\right)=\hat{I}\left(0_c,t\right)$, $V_J=\left({\Phi_0}/{2\pi}\right)\dot{\delta}+\left[\dot{\phi}(0_b)-\dot{\phi}(0_c)\right]$. Thus, under the low-excitation limit and rotating-wave approximation, i.e., the photon scattering is the desired elastic and any possibly created and annihilated of the photons at $x=0$ is neglected, we have
$$
\begin{aligned}
\hat{H}_{CBJJ-B}=& C_J{\hat{p}}_{\theta}\left[\dot{\phi}(0_b)-\dot{\phi}(0_c)\right]\\
=& i\hbar\sqrt{\frac{\kappa_l}{2\pi}}{\int {d \omega}} \left[a^{\dagger} l(\omega)-l^{\dagger}(\omega) a\right],
\end{aligned}
$$	
where $\kappa_l=Z_0 /4Z_J\, (l=b,c)$ describes the interaction between the CBJJ and the left/right traveling-wave photons, $Z_J=\sqrt{{L_J}/{C_J}}$ is the characteristic impedance of the Josephson junction. As a consequence, the Hamiltonian (with $\hbar=1$) of the system~\cite{Collett,Boutin,He}:
\begin{equation}
\begin{aligned}
H_B =& \left(\omega_{P}-\frac{i\gamma}{2}\right)a^{\dagger} a\\
&+\int d \omega\left[\omega b(\omega)^{\dagger} b(\omega)+i \sqrt{\frac{\kappa_{1}}{2 \pi}}\left(a^{\dagger} b(\omega)- a b(\omega)^{\dagger}\right)\right] \\
&+\int d \omega\left[\omega c(\omega)^{\dagger} c(\omega)+i \sqrt{\frac{\kappa_{2}}{2 \pi}}\left(a^{\dagger} c(\omega)-ac(\omega)^{\dagger} \right)\right],
\end{aligned}
\end{equation}
where $\gamma$ is decay rate of the cavity, $\kappa_1$ and $\kappa_2$ are the effective strengths of the boson coupled to the photons in the left and right sides of the transmission line, respectively. By using the standard input-output theory~\cite{He, Input-Output}, we get the relations:
\begin{equation}
\frac{d a}{d t}=\left(-i \omega_{p}-\frac{\kappa+\gamma}{2}\right) a+\sqrt{\kappa_{1}} b_{\rm{in}}(t)+\sqrt{\kappa_{2}} c_{\rm{in}}(t),
\end{equation}
and
\begin{equation}
\frac{d a}{d t}=\left(-i \omega_{p}+\frac{\kappa-\gamma}{2}\right) a
-\sqrt{\kappa_{1}} b_{\rm{out}}(t)-\sqrt{\kappa_{2}} c_{\rm{out}}(t),
\end{equation}
with $\kappa=\kappa_1+\kappa_2$,
$$
\hat{b}_{{\rm{in}}/{\rm{out}}}=\pm\frac{1}{\sqrt{2 \pi}} \int d \omega e^{-i \omega\left(t-t^{\prime}\right)} b_{0}(\omega),
$$
and
$$
\hat{c}_{{\rm{in}}/{\rm{out}}}=\pm\frac{1}{\sqrt{2 \pi}} \int d \omega e^{-i \omega\left(t-t^{\prime}\right)} c_{0}(\omega),
$$
are the input- and output fields, respectively.
After the Fourier transformation: $x(t)=\int_{-\infty}^{+\infty} e^{-i \omega\left(t-t_{0}\right)} x(\omega) d \omega/\sqrt{2 \pi}$, for $x(t)=a(t)$, $b_{\rm{in}}(t)$, $c_{\rm{in}}(t)$, $ b_{\rm{out}}(t)$, and $c_{\rm{out}}(t)$, respectively, we have
\begin{equation}
\begin{aligned}
-i \omega a(\omega)=& \left(-i \omega_{p}-\frac{\kappa-\gamma}{2}\right) a(\omega)\\
& +\sqrt{\kappa_{1}} b_{\rm{in}}(\omega)+\sqrt{\kappa_{2}} c_{\rm{in}}(\omega)
\end{aligned}
\end{equation}
and
\begin{equation}
\begin{aligned}
-i \omega a(\omega)=& \left(-i \omega_{p}+\frac{\kappa+\gamma}{2}\right) a(\omega)\\
& -\sqrt{\kappa_{1}} b_{\rm{out}}(\omega)-\sqrt{\kappa_{2}} c_{\rm{out}}(\omega).
\end{aligned}
\end{equation}
For the configuration shown in Fig.~\ref{Fig4:4}, we can assume that $c_{\rm{in}} (t)=0$. Consequently, we have
\begin{equation}
\begin{aligned}
&b_{\rm{in}}(\omega)+b_{\rm{out}}(\omega)=\sqrt{\kappa_{1}} a(\omega) \\
&c_{\rm{in}}(\omega)+c_{\rm{out}}(\omega)=\sqrt{\kappa_{2}} a(\omega).
\end{aligned}
\end{equation}
Therefore, the measurable transmitted- and phase shift spectra of the traveling-wave photons can be calculated as
\begin{equation}
T_{B}(\omega)=\left|\frac{c_{\rm{out}}(\omega)}{b_{\rm{in}}(\omega)}\right|^{2}=\frac{4 \kappa_{1} \kappa_{2}}{4\left(\omega-\omega_{p}\right)^{2}+(\kappa+\gamma)^{2}},
\end{equation}
and
\begin{equation}
\phi_{T_{B}}(\omega)=\arctan\left[-\frac{2\left(\omega-\omega_{p}\right)}{\kappa+\gamma}\right],
\end{equation}
respectively. In Fig.~4 we shows the spectra of the traveling-wave microwave photons scattered by a quantized CBJJ device with the typical parameters: $I_b=0, I_c=0.975{\rm{~\mu A}}$, $C=11.18{\rm{~pF}}$, and thus $\omega_p\sim 2\pi\times 2.595{\rm{~GHz}}$.
\begin{figure}[h]
\centering
 \subfigure{\includegraphics[width=4.2cm]{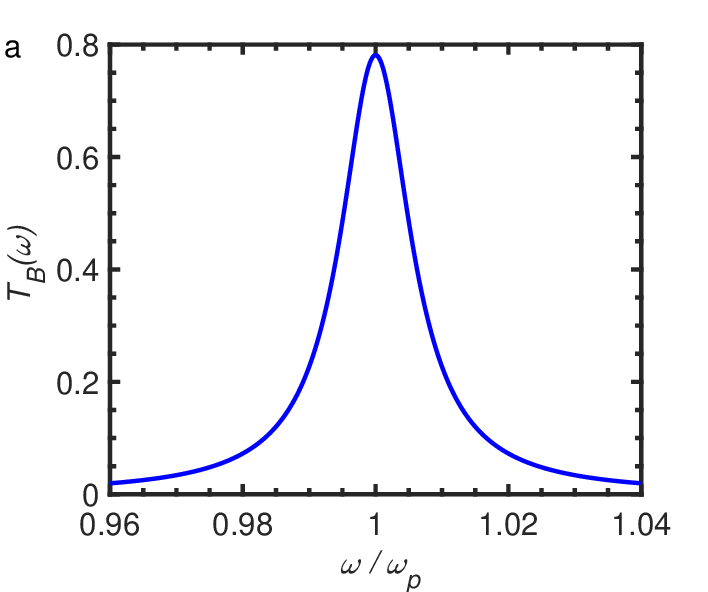}}
 \subfigure{\includegraphics[width=4.2cm]{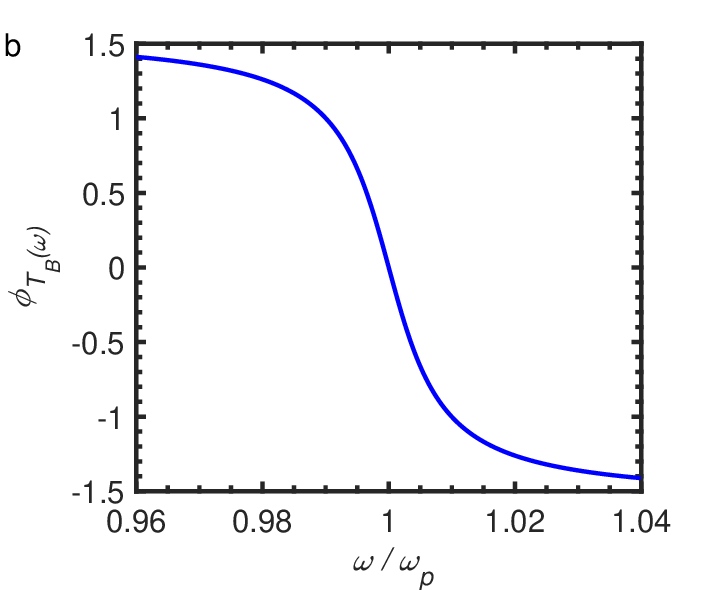}}
\caption{The transmitted spectrum (a) and phase shift spectrum (b) of the traveling-wave scattered by the CBJJ. Here, the relevant parameters are set as: $\kappa_1=0.004$, $\kappa_2=0.008$, $\gamma=0.0008$, and $\omega_{p}=2\pi\times 2.595{\rm{~GHz}}$.}
\label{Fig5:5}
\end{figure}
It is seen clearly that, if the CBJJ device works as a boson, the peak of the photon transmission is located as the eigenfrequency of the cavity, i.e., $\omega=\omega_p$, while the phase shift of the transmitted photon is zero.

Alternatively, if the CBJJ is biased near its critical current, e.g., $0.965<I_b/I_{c}\leq 0.98$, only two macroscopic bound state exist in the relevant local potential, yielding that the CBJJ works effectively as a two-level artificial atom, i.e., a fermion with the Hamilton $\hat{H}_{CBJJ}\sim\hat{H}_f$ shown in Eq.~5. Similarly, the validity of such an equivalence can be tested by probing the transport properties of the traveling-wave photons scattered by the device under the relevant bias. The generic model, shown schematically in Fig.~5, to describe the photon transport scattered by a two-level atom, i.e., a fermion, generated by the present CBJJ, can be described by the Hamiltonian~\cite{H-F1}:
\begin{equation}
\begin{aligned}
H_F= & \left(\omega_{01}-\frac{i\Gamma}{2}\right) \sigma_{z}\\
&+\int d \omega\left[\omega b(\omega)^{\dagger} b(\omega)+\sqrt{\frac{\eta_{1}}{2 \pi}}\left(\sigma_{+} b(\omega)-b(\omega)^{\dagger} \sigma_{-}\right)\right] \\
&+\int d \omega\left[\omega c(\omega)^{\dagger} c(\omega)+\sqrt{\frac{\eta_{2}}{2 \pi}}\left(\sigma_{+} c(\omega)-c(\omega)^{\dagger} \sigma_{-}\right)\right]
\end{aligned}
\end{equation}
with $\omega_{0}$ and $\omega_{1}$ being the eigenfrequencies of the bound states: $| 0\rangle$ and $| 1\rangle$, respectively. $\eta_{1,2}=\left({2\pi\lambda}/{\Phi_0}\right)^2 {Z_0}/{2\hbar}$ is the effective strength of the fermion coupled to the photon in the left/right transmission line.
\begin{figure}[h]
\centering
\includegraphics[width=6cm]{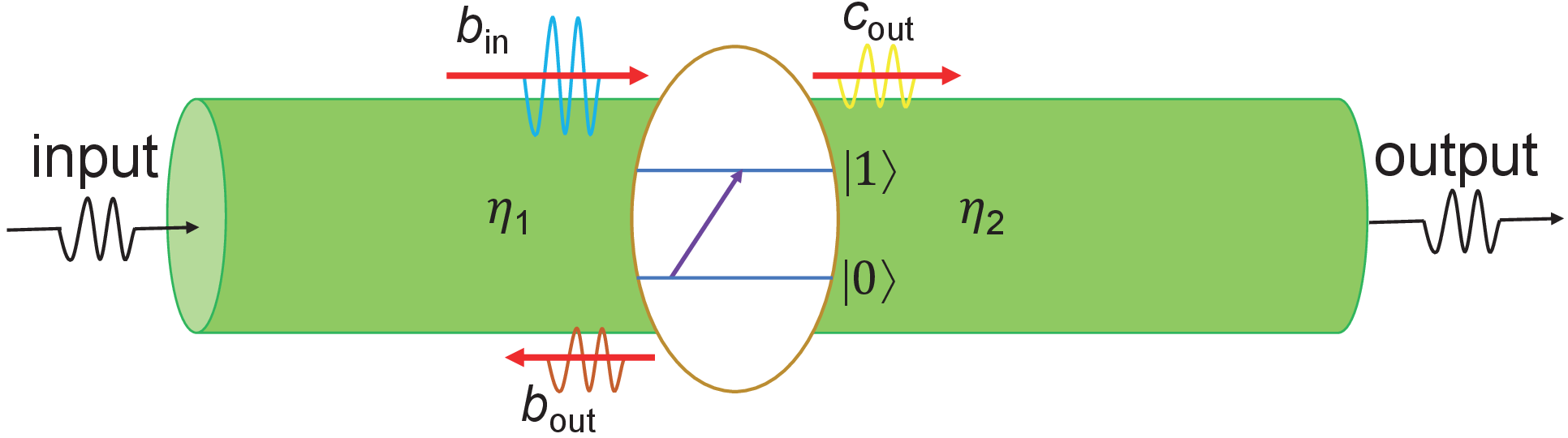}
\caption{A configuration of the traveling-wave photon scattered by a two-level atom, i.e., a fermion.}
\label{Fig6:6}
\end{figure}
It is easy to get the Heisenberg equations for the traveling-wave photon operators:
\begin{equation}
i\frac{d b(\omega)}{d t}=\omega b(\omega)+\sqrt{\frac{\eta_{1}}{2 \pi}}\sigma_{-}
\end{equation}
\begin{equation}
i\frac{d c(\omega)}{d t}=\omega c(\omega)+\sqrt{\frac{\eta_{2}}{2 \pi}}\sigma_{-}
\end{equation}
and the two-level atomic operator $\sigma_{-}$:
\begin{equation}
\begin{aligned}
i\frac{d\sigma_{-}}{d t}&=(\omega_{01}-\frac{i\gamma}{2})\sigma_{-}\\
&-\sqrt{\frac{\eta_{1}}{2 \pi}}\int d\omega\sigma_{z} b(\omega)-\sqrt{\frac{\eta_{2}}{2 \pi}}\int d\omega\sigma_{z} c(\omega)
\end{aligned}
\end{equation}
respectively. 
Defining again the in- and out operators:
$$
\hat{b}_{{\rm{in}}/{\rm{out}}}=\mp\frac{1}{\sqrt{2 \pi}} \int d \omega e^{-i \omega\left(t-t^{\prime}\right)} b_{0}(\omega),
$$
and
$$
\hat{c}_{{\rm{in}}/{\rm{out}}}=\mp\frac{1}{\sqrt{2 \pi}} \int d \omega e^{-i \omega\left(t-t^{\prime}\right)} c_{0}(\omega),
$$
of the traveling-wave photon scattered by the fermion and using again the standard input-output theory, we get the following input-output relations:
\begin{equation}
\begin{aligned}
\frac{d\sigma_{-}}{dt}&=(-i \omega_{01}-\frac{\eta+\gamma}{2}) \sigma_{-}(t) \\
&+i\sqrt{\eta_{1}}\sigma_{z} b_{\rm{in}}(t)+i\sqrt{\eta_{2}}\sigma_{z} c_{\rm{in}}(t)
\end{aligned}
\end{equation}
and
\begin{equation}
\begin{aligned}
\frac{d\sigma_{-}}{dt}&=(-i \omega_{01}+\frac{\eta-\gamma}{2}) \sigma_{-}(t)\\
&+i\sqrt{\eta_{1}}\sigma_{z} b_{\rm{out}}(t)+i\sqrt{\eta_{2}}\sigma_{z} c_{\rm{out}}(t)
\end{aligned}
\end{equation}
can be obtained. Consequently, we have
\begin{equation}
\begin{aligned}
&\sqrt{\eta_{1}}b_{\rm{in}}(t)-\sqrt{\eta_{2}}c_{\rm{out}}(t)=i\eta \sigma_{-}(t) \\
&\sqrt{\eta_{2}}c_{\rm{in}}(t)-\sqrt{\eta_{1}}b_{\rm{out}}(t)=i\eta \sigma_{-}(t)
\end{aligned}
\end{equation}
with $\eta=\eta_{1}+\eta_{2}$. Therefore, the scattering matrix $\langle k'|S|k\rangle$ between the input state $|k\rangle\sim b_{\rm{in}}^\dagger |0\rangle$ and the output state $|k'\rangle\sim b_{\rm{out}}^\dagger |0\rangle$ of the photons can be expressed as:
\begin{equation}
\begin{aligned}
\langle k'|S|k\rangle &=\frac{1}{\sqrt{2\pi}} \int dt\langle 0|c_{\rm{out}}(t)| k^{+}\rangle \\
&=t_{F}(k)\delta(k-p),
\end{aligned}
\end{equation}
yielding the amplitude of the photon transmission:
\begin{equation}
t_F(\omega)=\frac{t_{F}(k)+1}{2}=\frac{2(\omega-\omega_{01})+i\Gamma}{2(\omega-\omega_{01})+i(\eta+\Gamma)},
\end{equation}
specifically for $\eta_1=\eta_2$.
\begin{figure}[h]
 \centering
 \subfigure{\includegraphics[width=4.2cm]{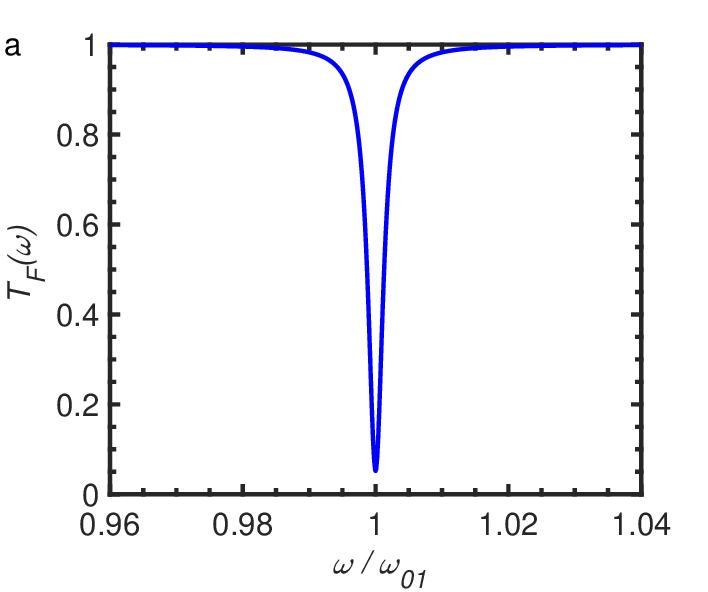}}
 \subfigure{\includegraphics[width=4.2cm]{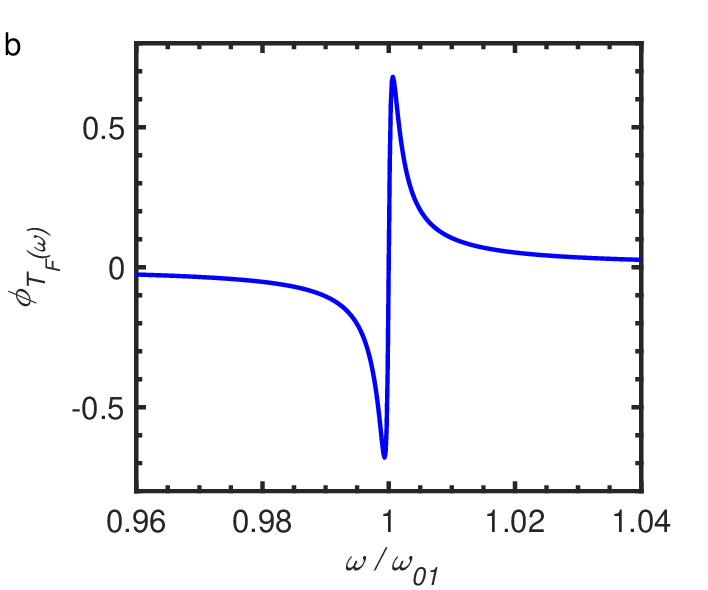}}
 \caption{The transmitted- (a) and phase shift (b) spectra of the travelling-wave photons scattered by a fermion generated by the CBJJ device. Here, the relevant parameters are set as: $\eta=0.0021$, $\Gamma=0.0062$, and $\omega_{01}=2\pi\times 2.42{\rm{~GHz}}$.}
 \label{Fig7:7}
\end{figure}
Obviously, the transmitted probability and phase shift of the traveling-wave photon can be expressed as
\begin{equation}
T_F(\omega)=|t_F(\omega)|^2=\frac{4(\omega-\omega_{01})^2+\gamma^2}{4(\omega-\omega_{01})^2+(\eta+\gamma)^2},
\end{equation}
and
\begin{equation}
\phi_{T_F(\omega)}=\arctan\left[-\frac{4(\omega-\omega_{01})^2+\gamma(\gamma+\eta)}{2\eta(\omega-\omega_{01})}\right],
\end{equation}
respectively. One can see from Fig.~6 that, the peak of the reflected photon takes place at $\omega=\omega_{01}$, i.e., the resonant photon is maximally reflected. Near such a resonant photon, the phase shift of the traveling-wave photon shows a manifestly critical behavior. These features are basically different from those of the boson scattering demonstrated above.

\section{Experiments and result analysis}

In this section, we report the experimental tests of the above arguments and the relevant theoretical predications, by fabricating the devices and measuring their microwave transport properties.

\subsection{Device fabrication and its physical parameter measurements}

By using the usual electron beam evaporations, the fabricated Josephson junction is embedded in a coplanar waveguide transmission line. Fig.~7 shows briefly the process for the fabrication of an $Al/AlO_x/Al$-junction. It mainly includes two steps; preparing a so-called ``Dolan bridge" by laser direct writing lithography, and fabricating the desired ``SIS" structure by the oblique deposition of aluminium films and oxidation.
\begin{figure}[h]
\centering
\includegraphics[width=3in,height=1in]{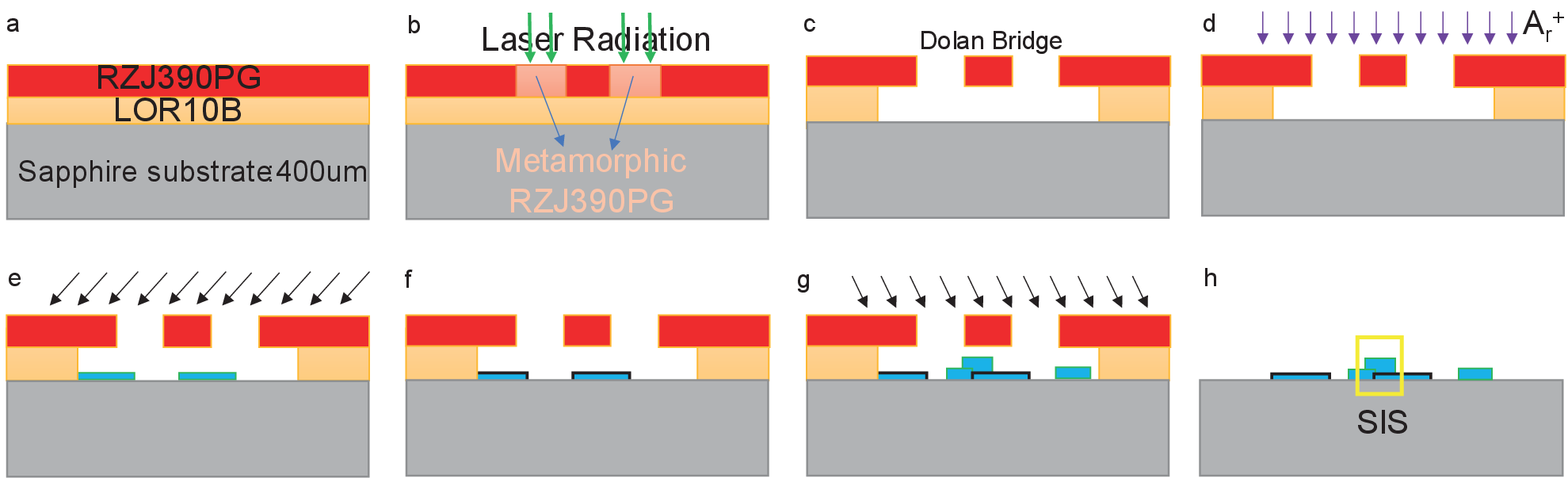}
\caption{A simplified process flow for the Josephson junction fabrication: (a) Uniform double layer photoresist  \, (b) Exposure \, (c) Development \, (d)ion etching oxide layer \, (e) Al by oblique evaporation \, (f) Oxidation \, (g) Al by positive rake angle evaporation \quad (h) Remove the photoresist. The yellow frame refers to the section of the junction area.}
\label{Fig8:8}
\end{figure}
The device was fabricated on a $400{\rm{~\mu m}}$ thick single-side polished sapphire substrate, which is cleaned with acetone and ultrasonic waves. The device structure is generated by using a laser direct-writing lithography process and then the wet etchings. The aluminum film is deposited onto the prepared ``Dolan Bridge"-structure on the sapphire substrate by the double-angle electron evaporations in the high vacuum environment. The thicknesses of the fabricated $Al/AlO_x/Al$-junction are estimated as: $80{\rm{~nm}}$ for the upper Al film, $1.2{\rm{~nm}}$ for the junction and $120{\rm{~nm}}$ for the lower Al film. As shown in Fig.~8
\begin{figure}[h]
\centering
\includegraphics[width=3in,height=2.2in]{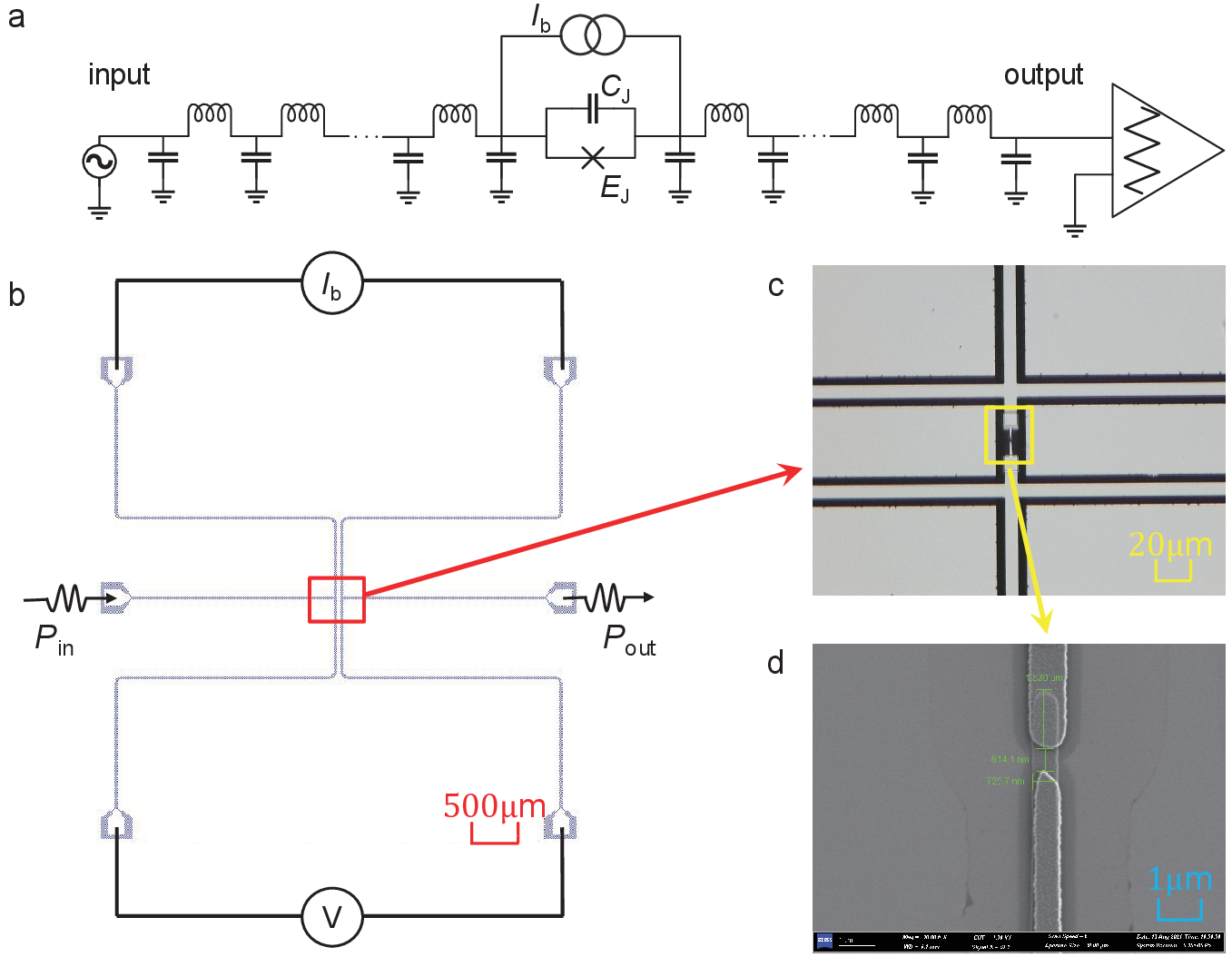}
\caption{The fabricated CBJJ device: (a) Its equivalent circuit; (b) The device structure, wherein the left wire connects the dc-current source, the right wire is used to measure the voltage across the junction, and the low-power traveling-wave microwave is driven along the middle transmission line; (c) A microscope image of the junction marked in the red box; and (d) A SEM image of the fabricated junction (in the orange box with the area being $1.26{\rm{~\mu m^2}}$).}
\label{Fig9:9}
\end{figure}
that, the widths of the transmission line and the gap are set as $\sim 10{\rm{~\mu m}}$, the area $A$ of the junction is $1.26{\rm{~\mu m^2}}$, respectively. Also, the distance between transmission line and the grounded one is set as $\sim 5{\rm{~\mu m}}$, to implement the $Z\thickapprox 50{\rm{~\Omega}}$ wave impedances matching with the external microwave devices.

To extract the physical parameters, typically such as the effective junction capacitance $C_J$, normal-state resistance $R_N$ and also the approximated critical current $I_c$, etc. of the fabricated Josephson junction, we first measured the I-V characteristic curve of the device by probing the switching currents of the junction, under the low-frequency ac drivings. A measurement circuit of the four-terminal method is shown in Fig.~9,
\begin{figure}[h]
\centering
\includegraphics[width=3in,height=2in]{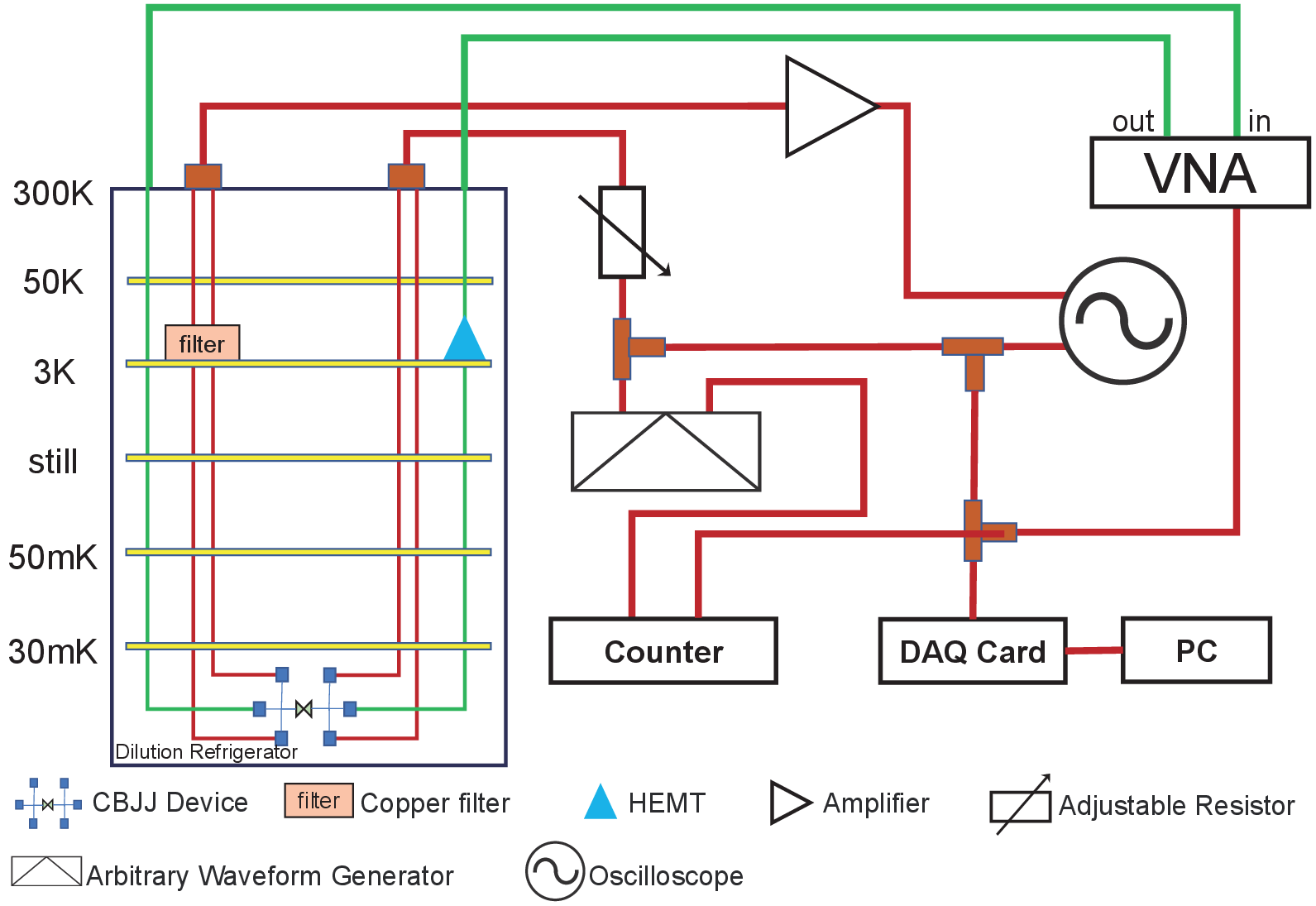}
\caption{The circuit to measure the I-V characteristic curve of the fabricated Josephson junction under the low-frequency AC driving.}
\label{Fig10:10}
\end{figure}
wherein the amplitude of the driving current is changed slowly and the voltage across the device is measured by a room temperature amplifier. A low-temperature copper filter and a large room-temperature resistor (whose resistance is much larger than the junction resistance $R_N$) are used to measure the slowly changed current through the junction. At $50{\rm{~mK}}$, the measured I-V curve of the fabricated Josephson junction is shown in Fig.~10(a), which shows that the junction works in the underdamped
\begin{figure}[h]
\centering
 \subfigure{\includegraphics[width=4.2cm]{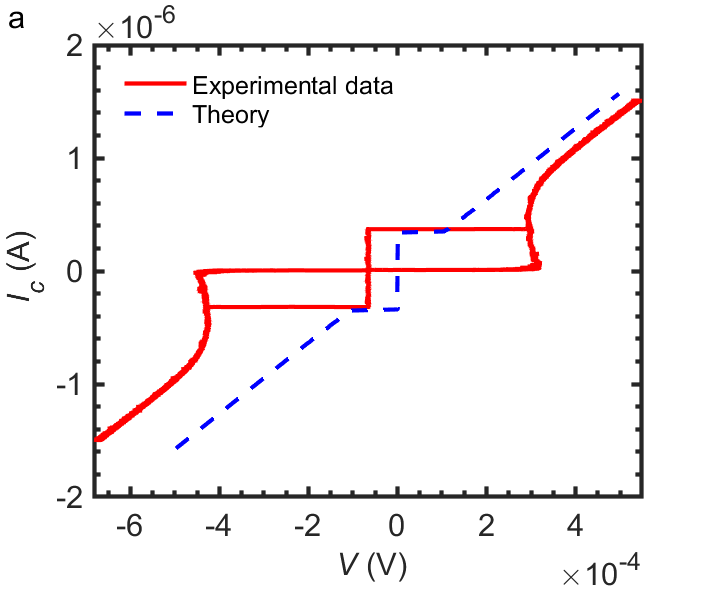}}
 \subfigure{\includegraphics[width=4.2cm]{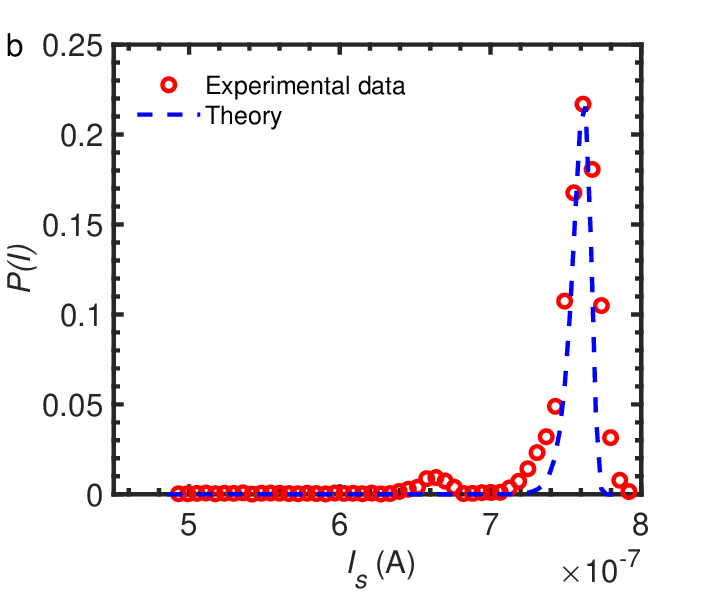}}
\caption{Low-frequency driving measurements of a fabricated Josephson junction. (a)The I-V curve, from which we know that the normal-state resistance of the junction is $R_{N}=290{\rm{~\Omega}}$; (b) The jump current measurements, wherein the junction is switched from the zero voltage state to the non-zero voltage state if the bias current is $I_s$.}
\label{Fig11:11}
\end{figure}
regime and its normal-state resistance is $R_{N}=290{\rm{~\Omega}}$. Consequently, with the superconducting gap of the Al: $\Delta=180{\rm{~\mu eV}}$, the critical current of the junction can be estimated roughly as $\tilde{I}_{c}=0.975{\rm{~\mu A}}$, according to the the Ambegaokar-Baratoff formula~\cite{ABF}: $\tilde{I}_c=\pi\Delta/(2eR_N)$.
On the other hand, the three-layer structure of SIS Junction can be treated as a parallel plate capacitor, as it is composed of two superconductors separated by a thin insulating layer. Therefore, the capacitance of the junction can be Phenomenologically calculated as: $\tilde{C}_{J}=\varepsilon_{r}\varepsilon_{0}A/d\approx 93{\rm{~fF}}$, with the area of the junction being $A \approx 1.26{\rm{~\mu m^2}}$, the thickness of the barrier layer being $d \approx 1.2{\rm{~nm}}$, and the vacuum permittivity and the relative permittivity of $AlO_x$ being $\varepsilon_0=10^{-12}{\rm{~F}}/{\rm{m}}$ and $\varepsilon_{r}=10{\rm{~F}}/{\rm{m}}$, respectively. However, as the junction is not an isolated device, it is realistically integrated with the other devices on chip, the effectively physical parameters should be determined by the relevant experimental measurements, and then simulated by the corresponding theoretical model such as the generic RCSJ model described by the following dynamical equation~\cite{RCSJ}:
\begin{equation}
I_b=\frac{\hbar C_J}{2e}\frac{d^{2}\delta}{dt^{2}}+\frac{\hbar}{2eR_N}\frac{d\delta}{dt}+I_{c}{\sin\delta},
\end{equation}
where $C_J$ and $I_c$ are the effective capacitance and critical current of the junction, $I_b$ is its low-frequency bias current (which is a sawtooth wave with the frequency of $f=71.3{\rm{~Hz}}$ and the amplitude of $I_{pp} = 3{\rm{~\mu A}}$, generated by an arbitrary waveform generator: Agilent 33250A), and $I_n$ the noise current (from typically such as the thermal fluctuation). First, the synchronization signal is triggered at the zero time as the start timing. Then, a slowly increasing bias current (with the slope being $dI/dt = 190{\rm{~\mu A}}/{\rm{~s}}$) is applied to the junction monitored by a voltage signal; when the voltage across the junction is switches from zero to a threshold non-zero value, the timing is ended and the bias current at this time is recorded as the jump current $I_s$. After repeated such a switch measurement with $N = 10000$ times, we get a statistical distribution of the measured jump currents $P(I_S)$, shown in Fig.~10(b).
It is seen that the maximal jump current is $\approx 0.78{\rm{~\mu A}}$, which is certainly less than the effective critical current of the junction. With the simulations based on the RCSJ model whose parameters are numerically adjustable, the experimentally measured statistical distribution of the jump current is well fitted by the red line in Fig.~10(b), from which the effective capacitance and critical current of the junction can be extracted as $C_J=93{\rm{~fF}}$ and $I_c=0.979{\rm{~\mu A}}$, respectively.

\subsection{Low-power microwave transport measurements}
With the fabricated CBJJ device whose physical parameters have been calibrated experimentally, we now verify the arguments predicted theoretically in Sec.~II by the microwave scattering measurements using the vector-network-analyzer (VNA) at 50mK.

First, under a sufficiently low dc-current bias, typically such as $I_b=0$, we had argued that the CBJJ device should behave as a boson-type macroscopic particle, i.e., a quantized linear harmonic oscillator. As a consequence, a peak of the measured transmitted spectrum of the applied traveling-wave microwave photons should be located at the plasma frequency $\omega_p$, at which the phase shift of the transmitted wave should be zero. It is clearly shown in Fig.~12 that the argument is true, i.e., the measured peak in the transmitted spectrum is revealed at $f_p=2.595{\rm{~GHz}}\sim\omega_p/2\pi$, where $\omega_p=16.305{\rm{~GHz}}$. The measured quality factor reads $Q_{I_b=0}=507$, which is approximated to the numerical simulation shown above. Also, the measured phase shift of the transmitted wave at the frequency $\omega=\omega_p$ is really zero.

\begin{figure}[h]
\centering
 \subfigure{\includegraphics[width=4.2cm]{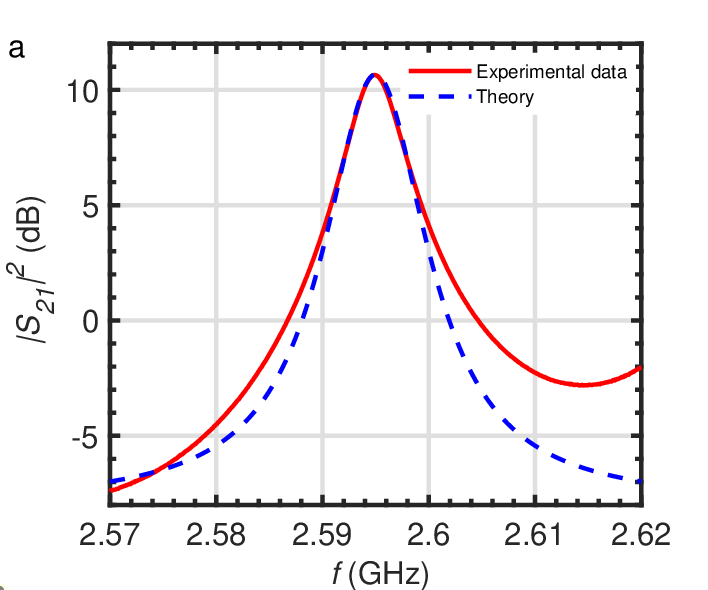}}
 \subfigure{\includegraphics[width=4.2cm]{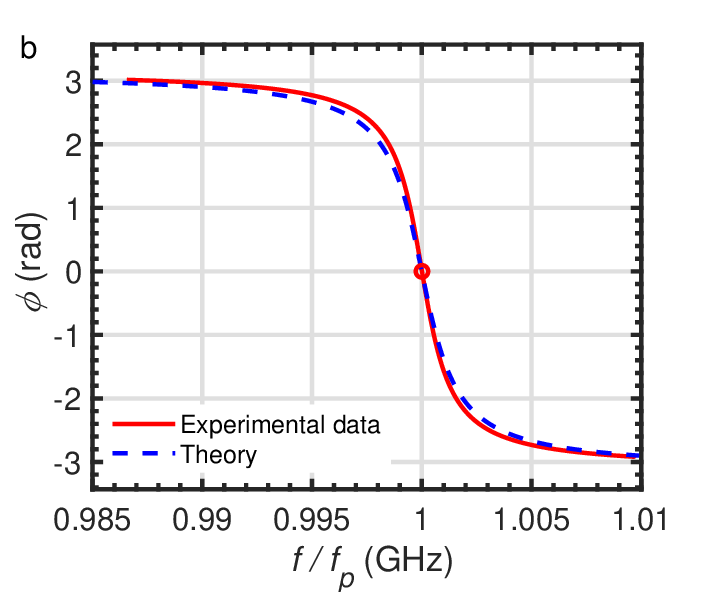}}
 \caption{The transmitted- and phase shift spectra of the traveling-wave photons scattered by the CBJJ without the dc-current bias; (a) The transmitted spectrum with the peak at $f_p=2.5949{\rm{~GHz}}$ and the quality factor $Q=507$. (b) The phase shift spectrum, in which the $\pi$-phase is shifted at $f=f_p$. Here, the red line represents the measured results and the blue lines are the fittings from the theoretical calculations. The power of the driving microwave is $P=-30{\rm{~dBm}}$ (i.e., under the low-excitation limit) and the fitted parameters are set as: $\kappa_1= 0.004$ and $\kappa_2=0.008$.}
\label{Fig12:12}
\end{figure}

Secondly, let us check experimentally if the quantized CBJJ device with the sufficiently large dc-current bias can be served as a fermion-type macroscopic particle. Typically, for the sufficiently large dc-current bias, such as $I_b=0.95 \mu A$ with $I_b/I_c=0.97$, Fig.~12 shows the relevant transmitted and phase shift spectra of the traveling-wave scattered by the device. It is seen clearly that, the spectra are very different from those for the device with low current biases. Here, the resonant photons are reflected completely and phase shift shows a transition behavior at the resonant point. This is the manifest feature of the traveling-wave photons scattered by a two-level atom (i.e., a fermion) predicted theoretically in the Sec.~II.
The fitted data show that the eigenfrequency of the fermion (i.e., the two-level artifical atom) is $f_{01}=\omega_{01}/(2 \pi)=2.4182{\rm{~GHz}}$ which is less significantly than $f_p$ observed previously. The effective junction capacitance is $C_{01}=2.55{\rm{~pF}}$, which is larger than $C_p$ estimated previously in the absence of the dc-current bias.
\begin{figure}[h]
\centering
 \subfigure{\includegraphics[width=4.2cm]{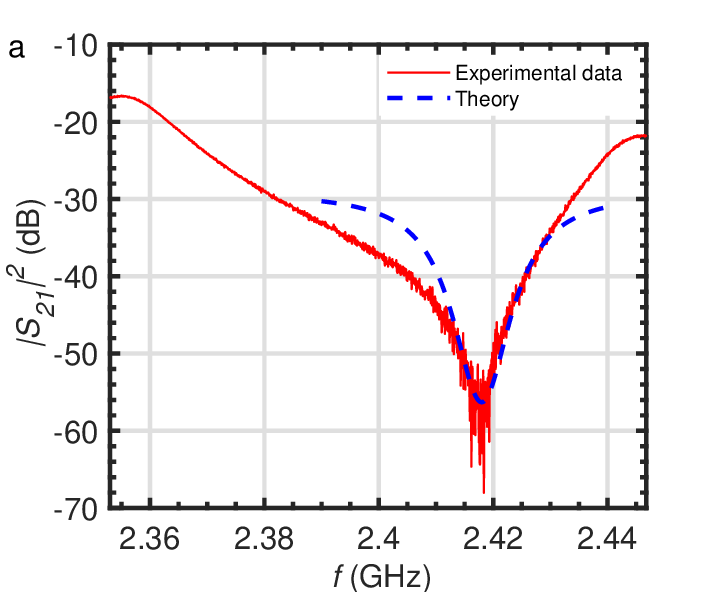}}
 \subfigure{\includegraphics[width=4.2cm]{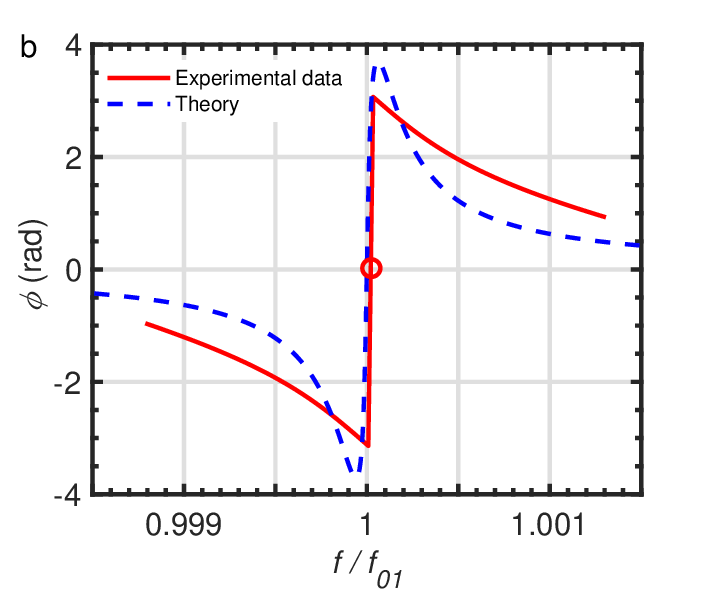}}
 \caption{The transmitted- and phase shift spectra of the traveling-wave photons scattered by the CBJJ with the dc-current bias $I_b=0.97 I_c$; (a) The transmitted spectrum, and (b) the phase shift spectrum.  Here, the power of the applied traveling wave is $P=-30{\rm{~dBm}}$,  the red lines represent the measured results and the blue lines are fitted by the  theoretical calculations with $\eta=0.005$ and $\gamma= 0.0012$.}
\label{Fig13:13}
\end{figure}
With the fitted data shown in Fig.~12, the critical current of the CBJJ without the current bias can be fitted as $I_c=0.979{\rm{~\mu A}}$, which is consistent with the one estimated by the measured I-V curve. The quality factor of the device can be observed  as $Q_b\sim 60$ for the significantly large dc-bias current. Also, the effective capacitance is calibrated as $C=11.18{\rm{~pF}}$ in this case, which is larger than the junction capacitance $C_J$ extracted by using the measured I-V curve, due to the effect of the parasitic capacitance under the high frequency drivings.

Immediately, beyond the extreme cases, i.e., the CBJJ works as either a boson under the bias $I_b=0$ or a fermion under te bias $I_b/I_c\leq 1$, the experimental measurements shown in Fig.~13 indicate that, with the increase of the bias dc-current, the frequency of the transmitted peak is shifted left and also their relative heights decrease with the lower qualities. Obviously, the stronger bosonic effect could be observed for the higher quality, if the CBJJ device is biased by the lower dc-current. The asymmetric effects observed here on the transmitted peaks and phase shifts around the resonant point are due to the practically existing impedance mismatch between the CBJJ and the transmission line. This implies that the boson-type %
\begin{figure}[h]
\centering
 \subfigure{\includegraphics[width=4.2cm]{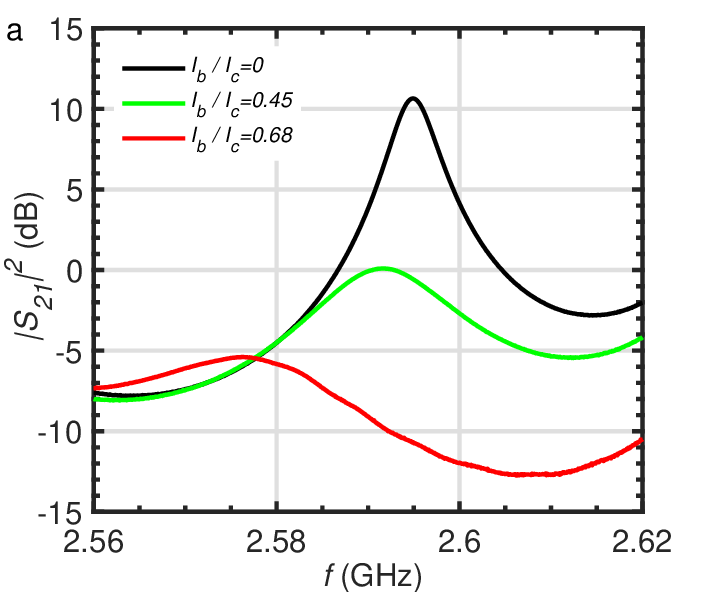}}
 \subfigure{\includegraphics[width=4.2cm]{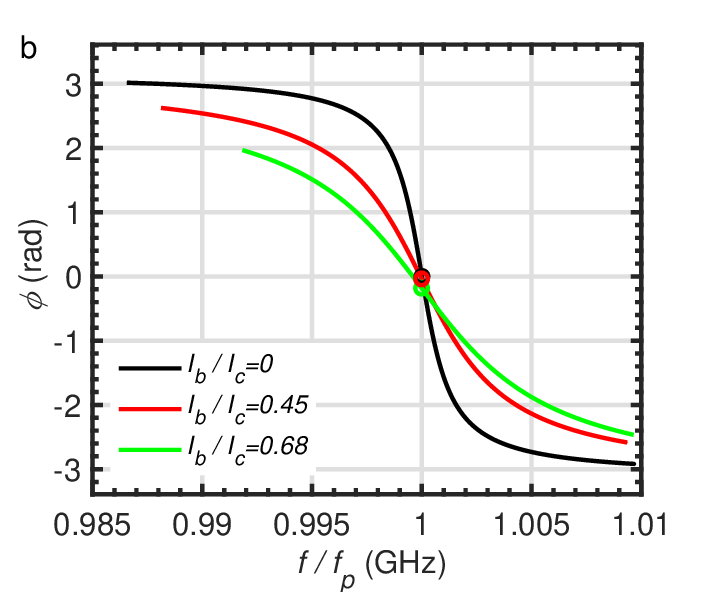}}
 \caption{The measured transmission of the traveling-wave photons scattered by the CBJJ with the dc-current biases being $I_b/I_c=0, 0.45$, and $0.65$, respectively: (a) The transmitted spectrum, and (b) the measured phase shift spectrum. Here, the drive power of the driving microwave is $P=-30{\rm{~dBm}}$.}
\label{Fig14:14}
\end{figure}
behavior is weakened. After that, if the biased current increases furthermore, then a series of transmitted dips would appear, until a single dip was observed. This process refers to the increase of the fermion-type behavior with the enhancing quality factor and decreased resonant eigenfrequency. Certainly, the effective junction capacitance is changed, if the CBJJ is biased by the different dc-currents and under the high frequency drivings.

\section{Conclusions and Discussions}
In a realistic three-dimensional space, the microscopic particles behave either the bosons satisfying the bosonic communication relation and fermions satisfying the ferimonic communication one, respectively. Usually, due to the self-average, the physical behavior of macroscopic object is classical, except in certain extreme environments such as the low-temperatures. In this work we first demonstrated experimentally that, under the ultra-low temperature and low-power microwave driving, a current-biased macroscopic Josephson junction can served as either a macroscopic ``boson" particle or a macroscopic ``fermion" one, depending on the amplitude of the biased dc-current. As a consequence, the CBJJ device can be utilized really in superconducting computing circuits to encode either the macroscopic qubit or work as the bosonic quantum data bus for coupling the distant qubits and reading out the information of the qubit(s).

Hopefully, the macroscopic devices demonstrated here with such the controllable transfer feature between the bosonic- and fermionic features provide the various novel applications for the simulations of the the quantum many-body physics~\cite{QMB-phys1,QMB-phys2,QMB-phys3}, microwave single-photon detection~\cite{MSPD1,MSPD2,MSPD3,MSPD4,MSPD5}, and also the salable quantum information processings~\cite{QIP1,QIP2,QIP3}, at macroscopic scale.

\section*{Acknowledgements}
This work is partially supported by the National Natural Science Foundation of China (NSFC) under Grant No.
11974290, and the National Key Research and Development Program of China (NKRDC) under Grant No. 2021YFA0718803.

\end{document}